\documentclass[traditabstract,letter]{aa}

\usepackage[displaymath, mathlines,switch]{lineno}

\usepackage{natbib,color}
\usepackage[normalem]{ulem}
\bibpunct{(}{)}{;}{a}{}{,} 

\usepackage{graphicx,hyperref,array,multirow}
\usepackage{mathrsfs,amssymb,amsmath}

\usepackage[varg]{txfonts}

\newcommand{\ud}{\mathrm{d}}
\newcommand{\be}{\begin{equation}}
\newcommand{\ee}{\end{equation}}

\newcommand*\xbar[1]{%
   \hbox{%
     \vbox{%
       \hrule height 0.5pt 
       \kern0.2ex
       \hbox{%
         \kern-0.1em
         \ensuremath{#1}%
         \kern-0.0em
       }%
     }%
   }%
}

\begin{document}

\title{Synchrotron emission in molecular cloud cores: the SKA view}

\author{Marco~Padovani and Daniele~Galli}

\authorrunning{M. Padovani and D. Galli}


\institute{INAF--Osservatorio Astrofisico di Arcetri, Largo E. Fermi 5, 50125
Firenze, Italy\\
\email{[padovani,galli]@arcetri.astro.it}
}


\abstract{Understanding the role of magnetic fields in star-forming regions is of fundamental importance.
In the near future, the exceptional sensitivity of SKA will offer a unique opportunity to evaluate 
the magnetic field strength 
in molecular clouds and cloud cores through synchrotron emission observations.
The most recent Voyager~1 data, together with Galactic synchrotron emission 
and Alpha Magnetic Spectrometer data, constrain
the flux of interstellar cosmic-ray electrons between $\approx3$~MeV and $\approx832$~GeV,
in particular in the energy range relevant for synchrotron emission in molecular cloud cores 
at SKA frequencies. 
Synchrotron radiation is entirely due to primary
cosmic-ray electrons, the relativistic flux of secondary leptons 
being completely negligible.
We explore the capability of SKA in detecting synchrotron emission in two starless molecular cloud cores
in the southern hemisphere, 
B68 and FeSt 1-457,
and we find that it will be possible to reach signal-to-noise
ratios of the order of 
$2-23$ at the lowest frequencies observable by SKA ($60-218$~MHz) 
with one hour of integration.
}


\maketitle

\section{Introduction}

During the last decades, many observational techniques have been developed to gather information on
the magnetic field strength and geometry in molecular clouds and star forming regions
such as Zeeman splitting of hyperfine molecular transitions \citep[e.g.][]{ct96}, optical and near-infrared polarisation of starlight \citep[e.g.][]{af08,aa11},
polarisation of sub-millimetre thermal dust emission \citep[e.g.][]{gb09,ag18}, maser emission polarisation \cite[e.g.][]{vh11},
Goldreich-Kylafis effect \citep{gk81},
and Faraday rotation \cite[e.g.][]{wr04}. Together, all these techniques contribute to elucidate the still controversial role
of magnetic fields in the process of star formation \citep[e.g.][]{mc99,mk04}.

An additional method to probe magnetic fields in molecular clouds 
is via synchrotron radiation produced by relativistic electrons braked by the cloud's magnetic fields 
\citep[e.g.][]{bm77}.  The intensity of the emission
depends only on the electron density per unit energy and the 
projection of the magnetic field on the plane perpendicular
to the line of sight.
However, this technique is constrained by two limitations: 
({\em i}\/) the poor knowledge of the interstellar (hereafter IS) flux of cosmic rays (hereafter CRs)
below $\approx500$~MeV and
({\em ii}\/) the limited sensitivity of current radiotelescopes. Today, 
thanks to the latest data release of the Voyager~1 spacecraft
\citep{cs16}, the flux of CR electrons down to about 3~MeV is well known. 
On the instrumental side, with the advent of the Square Kilometre Array (hereafter SKA), 
we are now approaching a new important era for high-resolution observations at radio frequencies.
As anticipated by \citet{db15}, in this paper we show that SKA will be able to detect synchrotron emission from molecular cloud
cores 
and provide  {significant} constraints on the magnetic field strength in molecular clouds.

It has been generally assumed that secondary
leptons 
by primary CR protons
are responsible for synchrotron emission in dense cores
\citep[e.g.][]{bm77,jp08}, 
while primary CR electrons are generally neglected (for an exception see \citealt{db15}). 
However, since H$_{2}$
column densities in dense cores do not exceed $\approx10^{23}$~cm$^{-2}$, one can confidently argue that 
synchrotron emission is dominated instead by primary CR electrons. In fact, as shown by \citet{pi18},
the flux of secondary leptons 
at relativistic energies is negligible 
(see their Fig.~7). Secondary leptons
dominate in the relativistic regime
only at column densities larger than $\approx10^{26}$~cm$^{-2}$, i.e. $\approx500$~g~cm$^{-2}$. 
At these very high column densities (typical of circumstellar discs, rather than molecular clouds), the bulk of relativistic 
secondary leptons is created by the decay of charged pions and pair production from photons produced by Bremsstrahlung and neutral pion decay,
see \citet{pi18}.

This paper is organised as follows: in Sect.~\ref{basiceqs} we recall the basic equations to compute
the synchrotron specific emissivity; in Sect.~\ref{elespec} we report the most recent determinations of the IS CR electron
flux; in Sect.~\ref{densmagprof} we describe the density and the magnetic field 
strength profiles that we use in
Sect.~\ref{applimod} to compute the synchrotron flux density of starless molecular cloud cores; 
in Sect.~\ref{discconc}
we discuss the implications for our results and summarise our most important findings.

\section{Basic equations}\label{basiceqs}
In this Section we recall the main equations to compute the expected radio emission of a molecular cloud core,
modelled as a spherically symmetric source.
Unless otherwise indicated, in the following we use cgs units.
The total power per unit frequency emitted by an electron of energy $E$ at frequency $\nu$ and radius $r$ is given by
\be\label{power}
P_{\nu}^{\rm em}(E,r)=\frac{\sqrt{3}e^{3}}{m_{e}c^{2}} B_{\perp}(r) F\left[\frac{\nu}{\nu_{c}(B_{\perp},E)}\right]\,,
\ee
where $e$ is the elementary charge, $m_{e}$ the electron mass, and $c$ the light speed
(see e.g. \citealt{longbook}).
Here, $B_{\perp}(r)$ is the projection of the magnetic field
on the plane perpendicular to the line of sight.
The function $F$ is defined by
\be
F(x)=x\int_{x}^{\infty}K_{5/3}(\xi)\ud\xi\,,
\ee
where $K_{5/3}$ is the modified Bessel function of order 5/3 and
\be\label{nuc}
\nu_{c}(B_{\perp},E)=
\frac{3eB_{\perp}}{4\pi m_{e}c}\left(\frac{E}{m_e c^2}\right)^{2}=
4.19\left(\frac{B_{\perp}}{\rm G}\right)%
\left(\frac{E}{m_e c^2}\right)^{2}~{\rm MHz}
\ee
is the frequency at which $F$ gets its maximum value.
The synchrotron specific emissivity $\epsilon_{\nu}(r)$ at frequency $\nu$ and radius $r$,
integrated over polarisations and assuming an isotropic
CR electron flux $j_{e}(E,r)$ (see Sect.~\ref{elespec}), namely the power per unit solid
angle and frequency produced within unit volume, is
\be\label{epsnu}
\epsilon_{\nu}(r) = \int_{m_{e}c^{2}}^{\infty}\frac{j_{e}(E,r)}{v_{e}(E)}P_{\nu}^{\rm em}(E,r)\,\ud E\,,
\ee
where 
$v_{e}(E)$ is the electron speed. 
The specific intensity is given by the specific emissivity integrated along a line of sight $s$.
For a spherical source of radius $R$ (see Sect.~\ref{densprof}), 
%
\be\label{Inu}
I_{\nu}(b) = 
\int_{0}^{s_{\rm max}(b)}\epsilon_{\nu}(s)\ud s=2\int_{b}^{R}%
\epsilon_{\nu}(r)\frac{r~\ud r}{\sqrt{r^{2}-b^{2}}}\,,
\ee
where $b$ is the impact parameter. 

In principle, if the synchrotron radiation is sufficiently strong, 
synchrotron self-absorption can take place. In this case the emitting electrons absorb synchrotron photons
and the emission is suppressed at low frequencies \citep[see e.g.][]{rl86}.
We verified that the expected synchrotron emission of prestellar cores with typical magnetic
field strengths of the order of $10~\mu{\rm G}-1~$mG is always optically thin even at the lowest 
frequencies 
reached by SKA.
Additional suppression of synchrotron emission can be produced by the Tsytovich-Razin effect, 
arising when relativistic electrons are surrounded by a plasma \citep[see e.g.][]{hw66,mabr13}. This effect is negligible
at frequencies $\nu\gg\nu_{\rm TR}=20(n_{e}/{\rm cm^{-3}})(B/{\rm G})^{-1}$, where $n_{e}$ is the electron volume density. 
Even maximising $\nu_{\rm TR}$ by assuming a magnetic field strength of 10~$\mu$G, a total volume density of
$10^{6}$~cm$^{-3}$, and a typical ionisation fraction of $10^{-7}$,
$\nu_{\rm TR}=200$~kHz. Therefore,  
we can safely neglect this effect since $\nu_{\rm TR}$ is below
the frequency range of 
SKA1-Low ($60-302$~MHz) and SKA1-Mid ($0.41-12.53$~GHz) according
to the most recent specifications\footnote{%
\url{https://astronomers.skatelescope.org/wp-content/uploads/2017/10/
SKA-TEL-SKO-0000818-01_SKA1_Science_Perform.pdf}.}.

\section{Interstellar cosmic-ray electrons}
\label{elespec}

\begin{figure}[!h]
\begin{center}
\resizebox{\hsize}{!}{\includegraphics{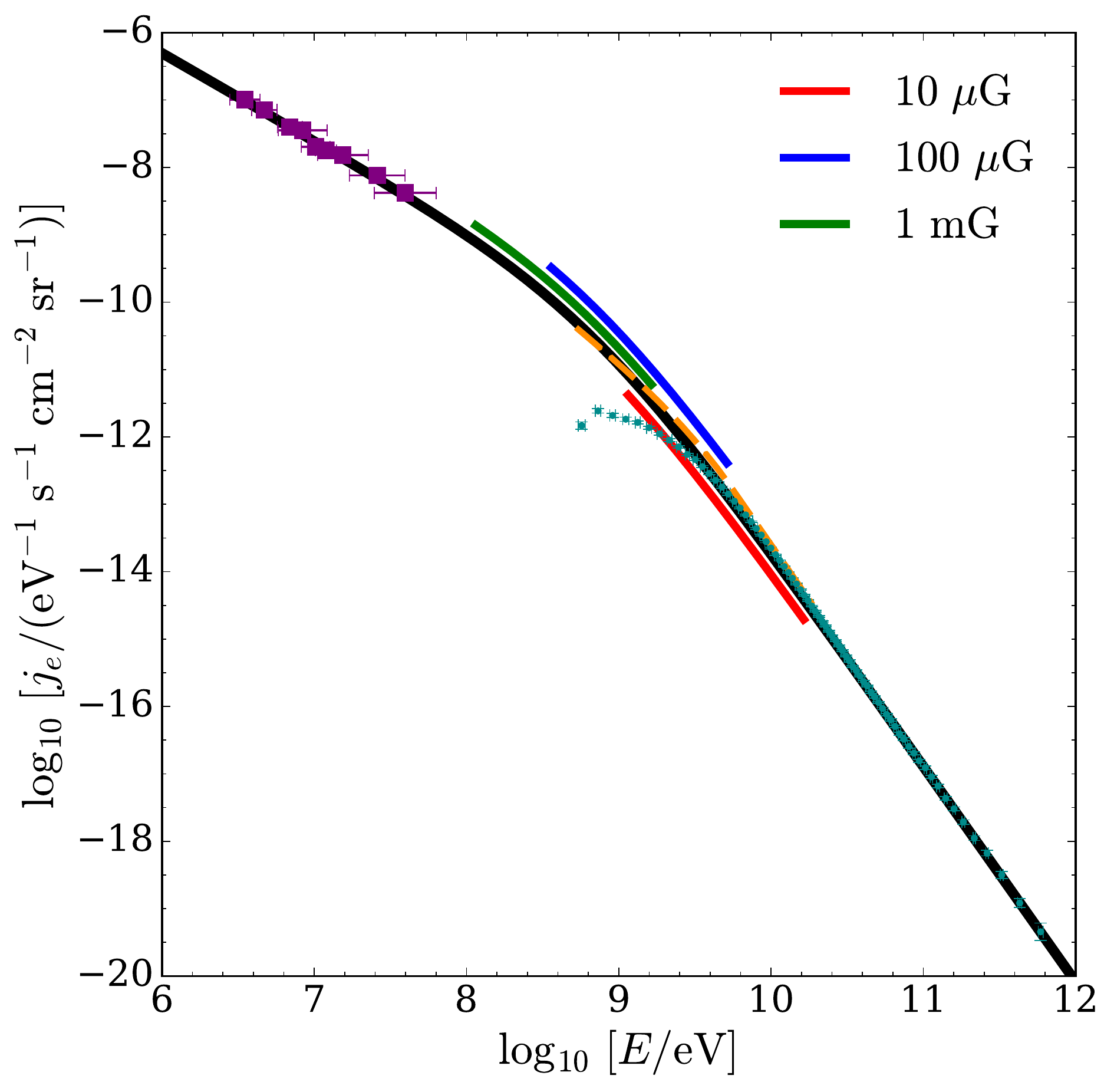}}
\caption{Flux of IS CR electrons ({\em black solid line}) as function of energy.
Data: Voyager~1~\citep[][{\em solid purple squares}]{cs16} at low energies,
AMS-02~\citep[][{\em solid cyan circles}]{aa14} at high energies.
The {\em red, blue,} and {\em green thick lines} show the energy range 
that mostly contribute to
synchrotron emission in the frequency range of SKA
for the values of the magnetic field strength listed in the legend (see Eq.~\ref{esyn}).
The {\em orange dashed line} shows the CR electron-positron flux obtained from Galactic
synchrotron emission \citep{o18}. 
} 
\label{Esyn_spectrum}
\end{center}
\end{figure}

The CR electron flux at energies below $\approx1$~GeV can be subject to propagation effects upon
their penetration into cores due to magnetohydrodynamical turbulence generated by CR protons,
causing modulation at low energies 
\citep{id18}. While the consequences of this process will be addressed in a following study (Ivlev et al., priv. comm.),
in this paper we assume the molecular cloud cores to be exposed to the IS CR electron flux
$j_{e}^{\rm IS}$. The latter is now well constrained 
between about 500~MeV and 20~GeV by Galactic synchrotron emission \citep{so11,o18} and
at lower energies ($3~{\rm MeV}\lesssim E\lesssim 40~{\rm MeV}$)
by the recent Voyager~1 observations \citep{cs16}. The latest Voyager data release shows that the electron flux 
increases with decreasing energy as $E^{-p}$ with a slope
$p=1.3$. At energies above a few GeV, the slope is constrained by Alpha Magnetic Spectrometer data
\citep[hereafter AMS-02;][]{aa14}, and is $p=3.2$.
\citet{ip15} and \citet{pi18} combined the low- and high-energy trends with the fitting formula
%
\be\label{jeism}
j_{e}^{\rm IS}(E)=j_{0}\frac{E^{a}}{(E+E_{0})^{b}}\,,
\ee
where $j_{0}=2.1\times10^{18}~\mathrm{eV^{-1}~s^{-1}~cm^{-2}~sr^{-1}}$, $a=-1.3$, $b=1.9$, $E_{0}=710$~MeV,
and $E$ is in eV.
Most of the synchrotron radiation is emitted around a frequency $\nu=0.29\nu_{c}$ \citep{longbook},
corresponding to
electrons of energy
\be\label{esyn}
E_{\rm syn}\approx464\left(\frac{\nu}{\rm Hz}\right)^{1/2}\left(\frac{B}{\rm G}\right)^{-1/2}~{\rm eV}\,. 
\ee
For typical values of the 
magnetic field strength 
measured in starless cores
(10~$\mu$G --1~mG), the corresponding values of $E_{\rm syn}$ for the frequency range
of SKA (see Sect.~\ref{basiceqs}) span the energy interval from
$\approx100$~MeV 
to $\approx20$~GeV as shown in Fig.~\ref{Esyn_spectrum}. 
For cloud cores with relatively strong magnetic fields (e.g. $\approx500~\mu$G) observed at 
$\nu\lesssim600$~MHz, $E_{\rm syn}$ is lower than $\approx500$~MeV. In this energy range Voyager data provide a better 
constraint on the CR electron flux than Galactic synchrotron emission and AMS-02 data.

\section{Density and magnetic field profiles}\label{densmagprof}

\subsection{Density profile}\label{densprof}
A common feature of starless cores derived from mm-continuum 
observations is a central flattening in the density profile, with typical maximum 
densities of the order of $10^{5}-10^{6}$~cm$^{-3}$, reaching $\approx2\times10^{7}$~cm$^{-3}$
in the starless core L1544 
\citep{kc10}. 
Various density profiles are used to reproduce observations of dense cores
such as the Bonnor-Ebert sphere or softened power laws. 
In the following we will adopt
the latter profile as given by \citet{tm02}
\be\label{denstaf}
n(r)=\frac{n_{0}}{1+(r/r_{0})^{q}}\,,
\ee
where $n_{0}$ is the central density, $r_{0}$ the radius of the central flat region,
and $q$ the asymptotic slope.

Since the specific emissivity is a function of radius
(Eq.~\ref{epsnu}), in principle one should consider the attenuation of the IS CR electron flux 
due to energy losses \citep{pgg09}. However, the typical maximum H$_{2}$
column densities of starless cores are of the order of $10^{23}$~cm$^{-2}$;
therefore 
only electrons with energies lower than 1~MeV can be stopped completely
\citep{pi18}. One should also take into account the fact that CRs follow helical trajectories along magnetic field lines, so 
that the effective column density passed through is larger than the corresponding line-of-sight column
density \citep[see e.g.][]{ph13}. Strong bending of field lines is not expected in starless cores: 
assuming a poloidal field configuration, the effective column density is larger at most by a factor of $\approx3$ 
(see Fig.~3 in \citealt{pg11}).
We can then assume the effective column density and the line-of-sight column density to be the same.
Besides, the minimum electron energy contributing to synchrotron emission for the range
of frequencies and magnetic field strengths considered here is $\approx100$~MeV,
corresponding to a 
value of the stopping range, of the order of $6\times10^{24}$~cm$^{-2}$.
Therefore, in Eq.~(\ref{epsnu}), we can safely neglect
the dependence of
the CR electron flux  
on column density
(i.e. on radius
$r$).


\subsection{Magnetic field strength profile}\label{magprof}

For the magnetic field strength we assume the relation
\be\label{magcru}
B(n)=B_{0}\left(\frac{n}{n_{0}}\right)^{\kappa}\,,
\ee
where $B_{0}$ is the value of the magnetic field strength at the central density $n_{0}$
and $\kappa\approx0.5-0.7$ is determined by observations \citep[see e.g.][]{cru12}.
We use Eq.~(\ref{magcru}) as a local relation to obtain the strength of the magnetic field 
from the radial profile of $n$.
While the choice of the density profile is unimportant for the attenuation of the CR electron flux
(see Sect.~\ref{densprof}),
it has important consequences on the 
profile of $B$ and, in turn, on 
the observed flux density.
In the following Section we compute the flux densities for different assumptions on the 
magnetic field 
profile.
Synchrotron emissivity depends on $B_{\perp}$
(see Eq.~\ref{epsnu}).
Here we simply assume $B_{\perp}=\pi B/4$ to account for a random orientation of the magnetic field 
with respect to the line of sight.

\section{Modelling}\label{applimod}

\subsection{General case}\label{generalcase}

\begin{figure}[!h]
\begin{center}
\resizebox{\hsize}{!}{\includegraphics{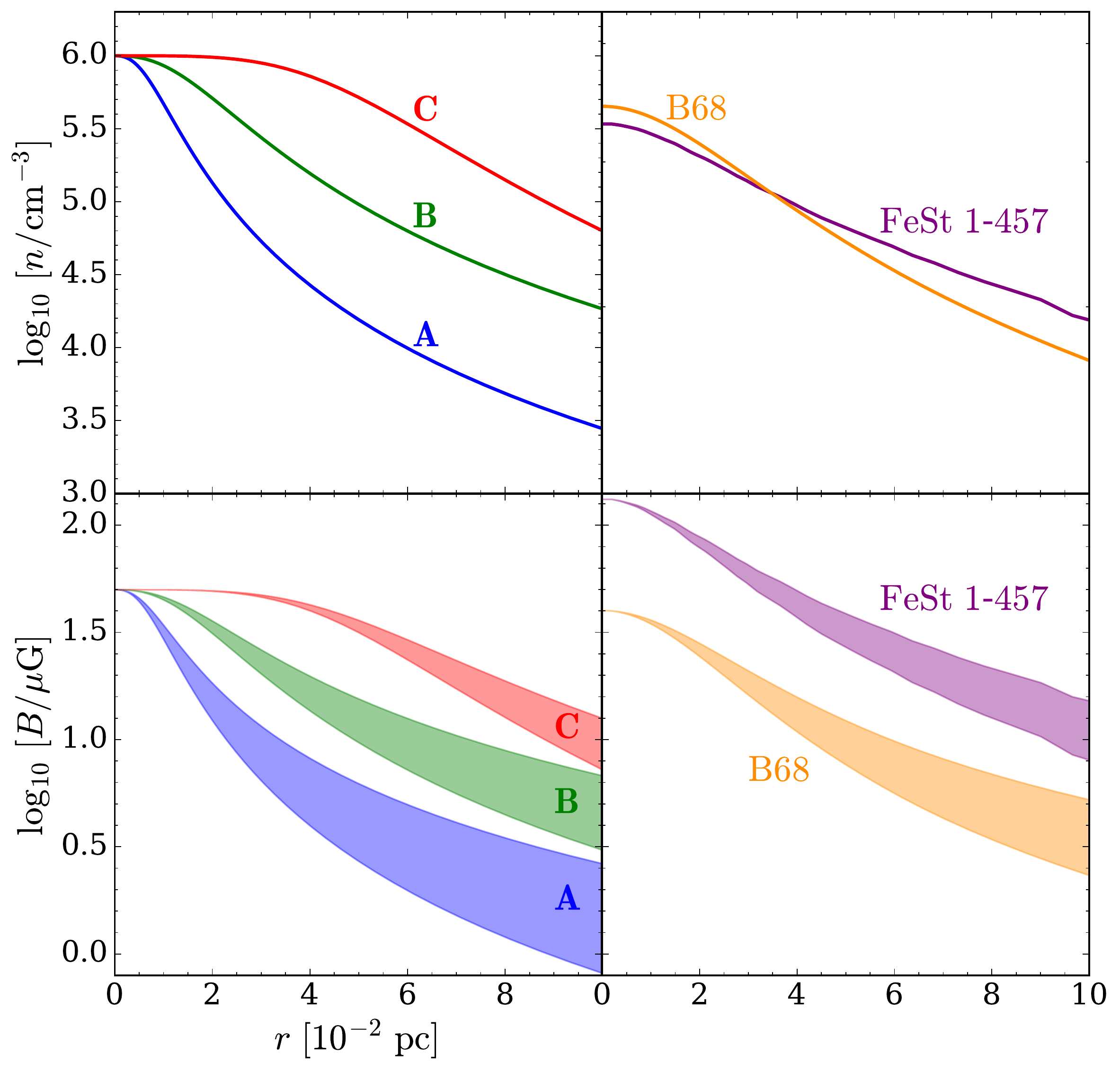}}
\caption{Density and magnetic field strength profiles as a function of radius (upper and lower panels,
respectively). 
Left column: models of starless cores described by
Eq.~(\ref{denstaf}), $r_{0}=14\arcsec, q=2.5$ (model A, {\em solid blue line}),
$r_{0}=30\arcsec, q=2.5$ (model B, {\em solid green line}), and
$r_{0}=75\arcsec, q=4$ (model C, {\em solid red line}).
Right column: starless cores FeSt 1-457 ({\em purple}) and B68 ({\em orange}).
Shaded areas in the lower panels 
encompass the curves obtained with Eq.~(\ref{magcru}) 
using $\kappa=0.5$ and 0.7 for models A, B, C and B68 and $\kappa=0.68$ and 0.88 for FeSt 1-457
(upper and lower boundary, respectively).
} 
\label{nBori}
\end{center}
\end{figure}

We consider
three models of starless cores with outer radius $R=0.1$~pc and central density $n_{0}=10^{6}$~cm$^{-3}$.
We vary the radius of the central flat region and the asymptotic slope (the parameters $r_{0}$ and $q$, respectively, 
in Eq.~\ref{denstaf}) ranging from a relatively peaked profile 
($r_{0}=14\arcsec,~q=2.5$, model A)
to an almost uniform profile ($r_{0}=75\arcsec,~q=4$, model C),
including an intermediate case ($r_{0}=30\arcsec,~q=2.5$, model B).
These profiles are similar to those used by~\citet{tm02} to describe low-mass starless cores in the Taurus 
molecular cloud.
The magnetic field strength is assumed equal 
to $B_{0}=50~\mu$G at
the core's centre in all models, and we consider the two cases 
$\kappa=0.5$ and 0.7~\citep{cru12}. The left column of Fig.~\ref{nBori} shows both the density and magnetic field strength profiles of our models.

\begin{figure*}[]
\begin{center}
\resizebox{\hsize}{!}{\includegraphics{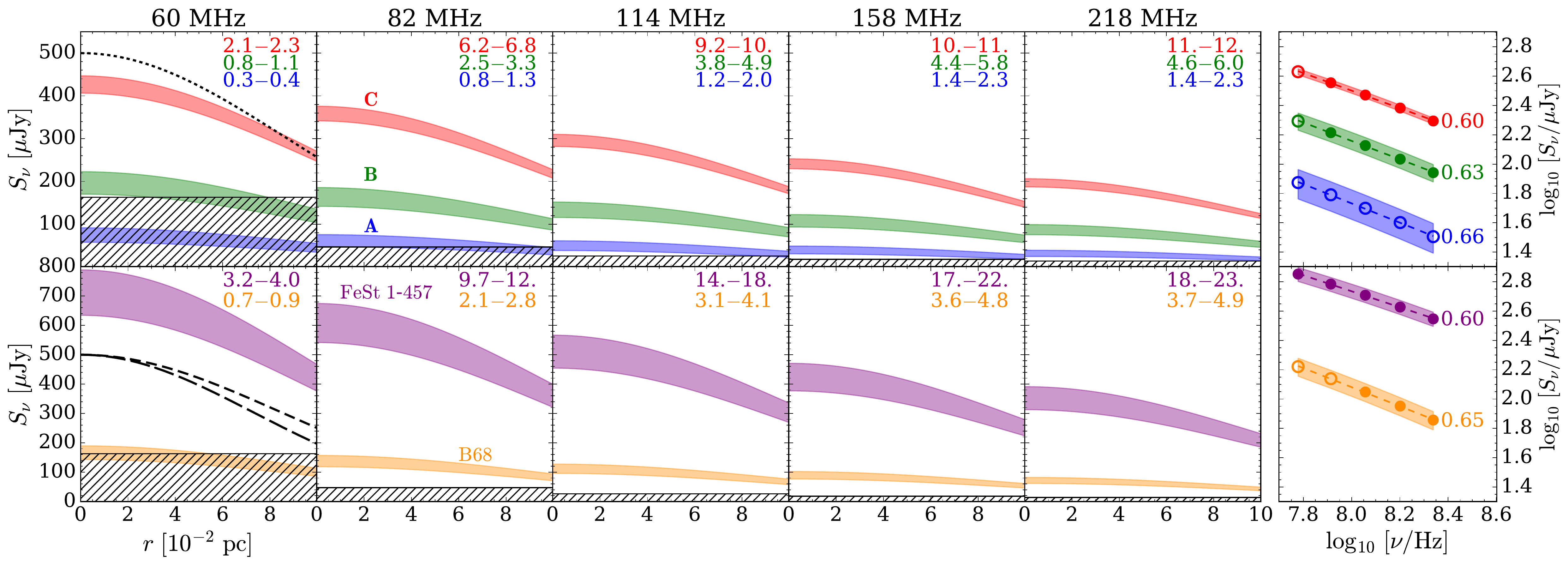}}
\caption{Radial flux density profiles  
for the starless core models described in
Sect.~\ref{generalcase} (upper row) and for B68 and FeSt 1-457
(lower row, see Sect.~\ref{particularcase}).
The observing frequency is shown in {\em black} at the top of each column,
while numbers in the upper-right corner of each subplot represent the 
radius-averaged S/N for the two values of $\kappa$
(0.5 and 0.7 for models A, B, C and B68,
and 0.68 and 0.88 for FeSt 1-457, see Eq.~\ref{magcru}).
Shaded areas encompass the curves obtained with Eq.~(\ref{magcru}) by using the two values of $\kappa$ 
(see Fig.~\ref{nBori} for colour-coding). The telescope beam is shown in the leftmost column for models A, B, and C
({\em dotted black line}, 300$\arcsec$), B68 ({\em short-dashed black line}, 330$\arcsec$),
and FeSt 1-457 ({\em long-dashed black line}, 284$\arcsec$).
{\em Hatched areas} display SKA sensitivities for one hour of integration
at different frequencies.
The two panels on the right side show the flux density as a function of frequency.
Empty (solid) circles refer to a S/N smaller (larger) than 3, respectively. 
The spectral index $\alpha$ is shown on the right of each curve.} 
\label{Snupaperfinal}
\end{center}
\end{figure*}

We compute the specific intensity from Eqs.~(\ref{epsnu}) and~(\ref{Inu}),
assuming the CR electron flux given by Eq.~(\ref{jeism}).
The observable quantity is the flux density $S\!_\nu$,
obtained by integrating over the solid angle
the specific intensity multiplied by the telescope pattern. The latter is assumed to be
a Gaussian with beam full width at half maximum $\theta_{b}=300\arcsec$
equal to the angular size of our starless core models at a distance of 140~pc
(non-resolved source).
%
The upper row of Fig.~\ref{Snupaperfinal} displays the 
flux density profiles at the five lowest frequencies of SKA1-Low
for the three models and the sensitivity limits for one hour of integration.
At higher frequencies a source of 300$\arcsec$ is resolved,
with consequent loss of flux. However, the total flux density can still be recovered by
convolution with a larger beam.

It is evident that what determines the maximum value of $S\!_\nu$
is not only the maximum 
magnetic field strength, but also the integrated value of $B$ along the line of sight
that, in turn, depends on the density profile:
the shallower the density profile, the higher is $S\!_{\nu}$. 
As a consequence, for one hour of integration, a centrally-peaked starless core such as model A is detectable
with a signal-to-noise (hereafter S/N) of about 2 only at frequencies higher than 158~MHz,
whereas a
less centrally concentrated core such as model B will be observable with ${\rm S/N}\gtrsim3$ at $\nu\gtrsim114$~MHz.
For an almost uniform density profile such as model C, a ${\rm S/N}\gtrsim2$ is promptly reached
at $\nu\gtrsim60$~MHz.

If the CR electron flux is proportional to $E^{-p}$, the synchrotron 
flux density varies as $S\!_{\nu}\propto\nu^{-\alpha}$,
where $\alpha=(p-1)/2$ is the spectral index \citep{rl86}. For our assumed CR electron flux,  $p\rightarrow1.3$ at low energies and $p\rightarrow3.2$ at high energies
(see Sect.~\ref{elespec}), therefore
$\alpha$ is expected to be intermediate between 0.2 and 1.1.  
In fact, our calculations 
show that $\alpha$ is about 0.6 
in the frequency range $60-218$~MHz
(see the two rightmost panels in Fig.~\ref{Snupaperfinal}).
Since in starless cores the IS CR electron flux is not attenuated, we expect the spectral
index $\alpha\approx0.6$ to be weakly dependent on density and magnetic field 
strength profiles at SKA1-Low frequencies as long as $10~\mu{\rm G}\lesssim B\lesssim1~{\rm mG}$. 

\begin{figure}[!h]
\begin{center}
\resizebox{.93\hsize}{!}{\includegraphics{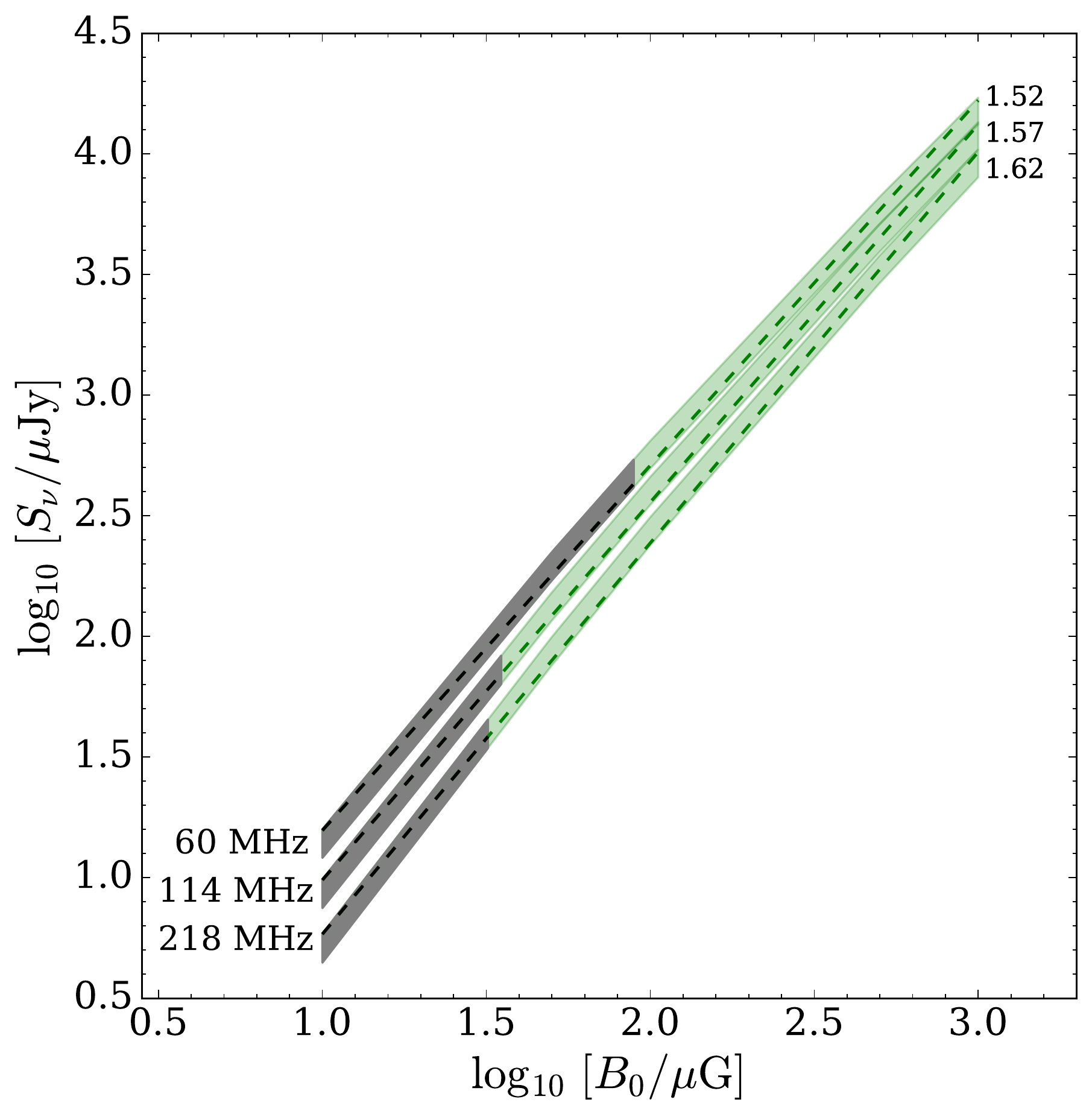}}
\caption{Flux density at 60, 114, and 218~MHz (lower-left labels) 
as a function of magnetic field strength $B_{0}$ for model B (see Sect.~\ref{generalcase}).
{\em Green shaded areas} encompass the curves obtained with Eq.~(\ref{magcru}) by using $\kappa=0.5$ and 0.7.
{\em Grey areas} correspond to ${\rm S/N}<3$ and dashed lines are power-law fits 
of $S\!_{\nu}\propto B_{0}^{\delta}$
with $\delta$ listed in the
upper-right corner.} 
\label{SvsB}
\end{center}
\end{figure}

The flux density depends weakly on the 
value of $\kappa$ (from 10\% in model C to
to 35\% in model A, for $\kappa$ ranging from 0.5 to 0.7), 
but it is strongly controlled by the strength of the magnetic field (Eq.~\ref{magcru}).
For a power-law electron flux proportional to $E^{-p}$ the synchrotron emissivity 
is $\epsilon_{\nu}\propto B^{\delta}$, where $\delta=(p+1)/2$ (see e.g. \citealt{rl86}).
Since $p$ varies between 1.3 and 3.2, 
we expect $S\!_{\nu}$
to increase with the magnetic field strength with a dependence intermediate
between $\delta=1.2$ and 2.1. 
Increasing the value of $B_{0}$ from 10~$\mu$G to 1~mG in the case of the core model B
we obtain the results shown in Fig.~\ref{SvsB} (models A and C give similar results). 
In the frequency range $60~{\rm MHz}-218~{\rm MHz}$
the flux density increases roughly in a power-law fashion with exponent $\approx1.5-1.6$.
As a consequence, a larger value of $B_{0}$ would make our modelled sources detectable 
with a shorter integration time.

Finally, we note that in this work we consider typical outer radii of the order of 0.1~pc, but for bigger starless cores --
especially with flat density profile -- the flux density would be correspondingly larger than that shown in 
Fig.~\ref{Snupaperfinal}.


\subsection{Magnetic field strength in FeSt 1-457 and B68}\label{particularcase}

In this Section we show the potential of SKA in observing two starless cores in 
the southern hemisphere: FeSt 1-457 (also known as Core 109 in the Pipe Nebula), 
and B68
in Ophiuchus.
The density profiles of FeSt 1-457 and B68 are taken from \citet{jg17} and \citet{gw02}, 
respectively. According to \cite{kt18}, the magnetic field at the centre of FeSt 1-457 is 
$B_{0}\approx132~\mu$G decreasing with a slope $\kappa=0.78\pm0.10$, while
the plane-of-sky magnetic field strength in B68 has been estimated by
\citet{kt09} as $\approx20~\mu$G. Since the total strength for a random field
is about twice that measured along any direction \citep{shubook99}, in the following
we consider $B_{0}=40~\mu$G for B68 
in Eq.~(\ref{magcru}).
The right column of Fig.~\ref{nBori} shows the density and the magnetic field strength profiles for these two
starless cores.

Similarly to the three modelled cores in Sect.~\ref{generalcase}, we compute the flux density 
using a beam size equal to the angular size of B68 and FeSt 1-457 (330$\arcsec$ and 284$\arcsec$, respectively%
\footnote{We assume a radius $R=0.1$~pc for both cores and a distance of 125~pc for B68 \citep{dd89}
and 145~pc for FeSt 1-457 \citep{af07}.}). The results are shown in the lower row of Fig.~\ref{Snupaperfinal}.
We find that 
a S/N larger than $\approx2$ is reached in one hour of integration 
at $\nu\gtrsim82$~MHz for B68, while FeSt 1-457 can be detected even at the lowest frequency $\nu=60$~MHz
with ${\rm S/N}\gtrsim3$.

\section{Conclusions}\label{discconc}

We explored the capability of SKA in detecting synchrotron emission from
starless cores. This observing 
facility
is very promising because it can provide 
robust constraints on the magnetic field strength, thus contributing to elucidate 
the role of magnetic fields in the star-formation process. Our main conclusions are:
\begin{itemize}
\item Thanks to the recent Voyager~1 measurements, combined with AMS-02 data and Galactic synchrotron emission, 
the flux of IS CR electrons is now well-known down to 3~MeV. 
Synchrotron emission is 
entirely dominated by primary CR electrons, while the contribution by secondary relativistic leptons
is negligible;
\item For typical values of the magnetic field strength in starless cores
($\approx10~\mu{\rm G}-1~{\rm mG}$), synchrotron emission in the SKA1-Low and SKA1-Mid frequency bands
is determined by electrons with energies between $\approx100$~MeV and $\approx20$~GeV. 
The stopping range at $\approx100$~MeV
is 
of the order of $6\times10^{24}$~cm$^{-2}$,
therefore 
the IS CR electron flux
contributing to synchrotron emission
is 
attenuated at 
H$_{2}$ column densities 
much in excess of the typical column densities of starless cores;
\item The synchrotron radiation emitted by a starless core 
depends on the integrated line-of-sight value of the magnetic field strength,
that, in turn, depends on the density profile. As a result, 
for the same maximum value of the magnetic field strength,
a starless core with a shallow density profile shows a larger synchrotron flux density than a
core with a peaked profile;
\item We modelled two starless cores in the southern hemisphere whose density profiles and typical 
magnetic field strengths are well-constrained by observations (B68 and FeSt 1-457)
and we found that SKA will be able to detect synchrotron emission at 
$82~{\rm MHz}\lesssim\nu\lesssim218~{\rm MHz}$ reaching a ${\rm S/N}\approx2-5$ and 
$\approx10-23$ in one hour of integration for B68 and FeSt 1-457, respectively;
FeSt 1-457 can be detected even at the lowest frequency $\nu=60$~MHz with a ${\rm S/N}\gtrsim3$;
\item We found a typical spectral index $\alpha=(p-1)/2\approx0.6$ 
intermediate between the values 
0.2 and 1.1 resulting from the low- and high-energy asymptotic slopes
of the primary CR electron flux ($p\rightarrow1.3$ and $p\rightarrow3.2$, respectively). 
Since in starless cores the IS CR electron flux is not attenuated by losses
(although it can be modulated by self-generated turbulence, see~\citealt{id18}), we expect $\alpha\approx0.6$
to be weakly dependent on density and magnetic field strength profiles at SKA1-Low frequencies.
\end{itemize}

Our study shows that SKA is a powerful instrument to constrain the strength of magnetic
fields in molecular cloud cores. This facility and the proposed next generation Very Large Array \citep[ngVLA,][]{mb18} 
have the capabilities to shed light in the near future on the role of magnetic fields in the process of
star formation.

\begin{acknowledgements} 
MP acknowledges funding from the European Unions Horizon 2020 research and innovation programme under the Marie Sk\l{}odowska-Curie grant agreement No 664931. The authors thank Maite Beltr\'an, Riccardo Cesaroni, and Francesco
Fontani 
for valuable discussions.
\end{acknowledgements}


\begin{thebibliography}{38}
\expandafter\ifx\csname natexlab\endcsname\relax\def\natexlab#1{#1}\fi

\bibitem[{{Aguilar} {et~al.}(2014){Aguilar}, {Aisa}, {Alvino}, {Ambrosi},
  {Andeen}, {Arruda}, {Attig}, {Azzarello}, {Bachlechner}, {Barao}, \&
  et~al.}]{aa14}
{Aguilar}, M., {Aisa}, D., {Alvino}, A., {et~al.} 2014, Physical Review
  Letters, 113, 121102

\bibitem[{{Alves} {et~al.}(2011){Alves}, {Acosta-Pulido}, {Girart}, {Franco},
  \& {L{\'o}pez}}]{aa11}
{Alves}, F.~O., {Acosta-Pulido}, J.~A., {Girart}, J.~M., {Franco}, G.~A.~P., \&
  {L{\'o}pez}, R. 2011, \aj, 142, 33

\bibitem[{{Alves} \& {Franco}(2007)}]{af07}
{Alves}, F.~O. \& {Franco}, G.~A.~P. 2007, \aap, 470, 597

\bibitem[{{Alves} {et~al.}(2008){Alves}, {Franco}, \& {Girart}}]{af08}
{Alves}, F.~O., {Franco}, G.~A.~P., \& {Girart}, J.~M. 2008, \aap, 486, L13

\bibitem[{{Alves} {et~al.}(2018){Alves}, {Girart}, {Padovani}, {Galli},
  {Franco}, {Caselli}, {Vlemmings}, {Zhang}, \& {Wiesemeyer}}]{ag18}
{Alves}, F.~O., {Girart}, J.~M., {Padovani}, M., {et~al.} 2018, \aap, 616, A56

\bibitem[{{Brown} \& {Marscher}(1977)}]{bm77}
{Brown}, R.~L. \& {Marscher}, A.~P. 1977, \apj, 212, 659

\bibitem[{{Crutcher}(2012)}]{cru12}
{Crutcher}, R.~M. 2012, \araa, 50, 29

\bibitem[{{Crutcher} {et~al.}(1996){Crutcher}, {Troland}, {Lazareff}, \&
  {Kazes}}]{ct96}
{Crutcher}, R.~M., {Troland}, T.~H., {Lazareff}, B., \& {Kazes}, I. 1996, \apj,
  456, 217

\bibitem[{{Cummings} {et~al.}(2016){Cummings}, {Stone}, {Heikkila}, {Lal},
  {Webber}, {J{\'o}hannesson}, {Moskalenko}, {Orlando}, \& {Porter}}]{cs16}
{Cummings}, A.~C., {Stone}, E.~C., {Heikkila}, B.~C., {et~al.} 2016, \apj, 831,
  18

\bibitem[{{de Geus} {et~al.}(1989){de Geus}, {de Zeeuw}, \& {Lub}}]{dd89}
{de Geus}, E.~J., {de Zeeuw}, P.~T., \& {Lub}, J. 1989, \aap, 216, 44

\bibitem[{{Dickinson} {et~al.}(2015){Dickinson}, {Beck}, {Crocker}, {Crutcher},
  {Davies}, {Ferri{\`e}re}, {Fuller}, {Jaffe}, {Jones}, {Leahy}, {Murphy},
  {Peel}, {Orlando}, {Porter}, {Protheroe}, {Strong}, {Robishaw}, {Watson}, \&
  {Yusef-Zadeh}}]{db15}
{Dickinson}, C., {Beck}, R., {Crocker}, R., {et~al.} 2015, Advancing
  Astrophysics with the Square Kilometre Array (AASKA14), 102

\bibitem[{{Galli} {et~al.}(2002){Galli}, {Walmsley}, \& {Gon{\c
  c}alves}}]{gw02}
{Galli}, D., {Walmsley}, M., \& {Gon{\c c}alves}, J. 2002, \aap, 394, 275

\bibitem[{{Girart} {et~al.}(2009){Girart}, {Beltr{\'a}n}, {Zhang}, {Rao}, \&
  {Estalella}}]{gb09}
{Girart}, J.~M., {Beltr{\'a}n}, M.~T., {Zhang}, Q., {Rao}, R., \& {Estalella},
  R. 2009, Science, 324, 1408

\bibitem[{{Goldreich} \& {Kylafis}(1981)}]{gk81}
{Goldreich}, P. \& {Kylafis}, N.~D. 1981, \apjl, 243, L75

\bibitem[{{Hornby} \& {Williams}(1966)}]{hw66}
{Hornby}, J.~M. \& {Williams}, P.~J.~S. 1966, \mnras, 131, 237

\bibitem[{{Ivlev} {et~al.}(2018){Ivlev}, {Dogiel}, {Chernyshov}, {Caselli},
  {Ko}, \& {Cheng}}]{id18}
{Ivlev}, A.~V., {Dogiel}, V.~A., {Chernyshov}, D.~O., {et~al.} 2018, \apj, 855,
  23

\bibitem[{{Ivlev} {et~al.}(2015){Ivlev}, {Padovani}, {Galli}, \&
  {Caselli}}]{ip15}
{Ivlev}, A.~V., {Padovani}, M., {Galli}, D., \& {Caselli}, P. 2015, \apj, 812,
  135

\bibitem[{{Jones} {et~al.}(2008){Jones}, {Protheroe}, \& {Crocker}}]{jp08}
{Jones}, D.~I., {Protheroe}, R.~J., \& {Crocker}, R.~M. 2008, \pasa, 25, 161

\bibitem[{{Ju{\'a}rez} {et~al.}(2017){Ju{\'a}rez}, {Girart}, {Frau}, {Palau},
  {Estalella}, {Morata}, {Alves}, {Beltr{\'a}n}, \& {Padovani}}]{jg17}
{Ju{\'a}rez}, C., {Girart}, J.~M., {Frau}, P., {et~al.} 2017, \aap, 597, A74

\bibitem[{{Kandori} {et~al.}(2009){Kandori}, {Tamura}, {Tatematsu}, {Kusakabe},
  {Nakajima}, \& {Nakajima}}]{kt09}
{Kandori}, R., {Tamura}, M., {Tatematsu}, K.-i., {et~al.} 2009, in IAU
  Symposium, Vol. 259, Cosmic Magnetic Fields: From Planets, to Stars and
  Galaxies, ed. K.~G. {Strassmeier}, A.~G. {Kosovichev}, \& J.~E. {Beckman},
  107--108

\bibitem[{{Kandori} {et~al.}(2018){Kandori}, {Tomisaka}, {Tamura}, {Saito},
  {Kusakabe}, {Nakajima}, {Kwon}, {Nagayama}, {Nagata}, \& {Tatematsu}}]{kt18}
{Kandori}, R., {Tomisaka}, K., {Tamura}, M., {et~al.} 2018, \apj, 865, 121

\bibitem[{{Keto} \& {Caselli}(2010)}]{kc10}
{Keto}, E. \& {Caselli}, P. 2010, \mnras, 402, 1625

\bibitem[{{Longair}(2011)}]{longbook}
{Longair}, M.~S. 2011, {High Energy Astrophysics}

\bibitem[{{Mac Low} \& {Klessen}(2004)}]{mk04}
{Mac Low}, M.-M. \& {Klessen}, R.~S. 2004, Reviews of Modern Physics, 76, 125

\bibitem[{{Mouschovias} \& {Ciolek}(1999)}]{mc99}
{Mouschovias}, T.~C. \& {Ciolek}, G.~E. 1999, in NATO Advanced Science
  Institutes (ASI) Series C, Vol. 540, NATO Advanced Science Institutes (ASI)
  Series C, ed. C.~J. {Lada} \& N.~D. {Kylafis}, 305

\bibitem[{{Munar-Adrover} {et~al.}(2013){Munar-Adrover}, {Bosch-Ramon},
  {Paredes}, \& {Iwasawa}}]{mabr13}
{Munar-Adrover}, P., {Bosch-Ramon}, V., {Paredes}, J.~M., \& {Iwasawa}, K.
  2013, \aap, 559, A13

\bibitem[{{Murphy} {et~al.}(2018){Murphy}, {Bolatto}, {Chatterjee}, {Casey},
  {Chomiuk}, {Dale}, {de Pater}, {Dickinson}, {Di Francesco}, {Hallinan},
  {Isella}, {Kohno}, {Kulkarni}, {Lang}, {Lazio}, {Leroy}, {Loinard},
  {Maccarone}, {Matthews}, {Osten}, {Reid}, {Riechers}, {Sakai}, {Walter}, \&
  {Wilner}}]{mb18}
{Murphy}, E.~J., {Bolatto}, A., {Chatterjee}, S., {et~al.} 2018, ArXiv e-prints
  [\eprint[arXiv]{1810.07524}]

\bibitem[{{Orlando}(2018)}]{o18}
{Orlando}, E. 2018, \mnras, 475, 2724

\bibitem[{{Padovani} \& {Galli}(2011)}]{pg11}
{Padovani}, M. \& {Galli}, D. 2011, \aap, 530, A109

\bibitem[{{Padovani} {et~al.}(2009){Padovani}, {Galli}, \& {Glassgold}}]{pgg09}
{Padovani}, M., {Galli}, D., \& {Glassgold}, A.~E. 2009, \aap, 501, 619

\bibitem[{{Padovani} {et~al.}(2013){Padovani}, {Hennebelle}, \& {Galli}}]{ph13}
{Padovani}, M., {Hennebelle}, P., \& {Galli}, D. 2013, \aap, 560, A114

\bibitem[{{Padovani} {et~al.}(2018){Padovani}, {Ivlev}, {Galli}, \&
  {Caselli}}]{pi18}
{Padovani}, M., {Ivlev}, A.~V., {Galli}, D., \& {Caselli}, P. 2018, \aap, 614,
  A111

\bibitem[{{Rybicki} \& {Lightman}(1986)}]{rl86}
{Rybicki}, G.~B. \& {Lightman}, A.~P. 1986, {Radiative Processes in
  Astrophysics}, 400

\bibitem[{{Shu} {et~al.}(1999){Shu}, {Allen}, {Shang}, {Ostriker}, \&
  {Li}}]{shubook99}
{Shu}, F.~H., {Allen}, A., {Shang}, H., {Ostriker}, E.~C., \& {Li}, Z.-Y. 1999,
  in NATO Advanced Science Institutes (ASI) Series C, Vol. 540, NATO Advanced
  Science Institutes (ASI) Series C, ed. C.~J. {Lada} \& N.~D. {Kylafis}, 193

\bibitem[{{Strong} {et~al.}(2011){Strong}, {Orlando}, \& {Jaffe}}]{so11}
{Strong}, A.~W., {Orlando}, E., \& {Jaffe}, T.~R. 2011, \aap, 534, A54

\bibitem[{{Tafalla} {et~al.}(2002){Tafalla}, {Myers}, {Caselli}, {Walmsley}, \&
  {Comito}}]{tm02}
{Tafalla}, M., {Myers}, P.~C., {Caselli}, P., {Walmsley}, C.~M., \& {Comito},
  C. 2002, \apj, 569, 815

\bibitem[{{Vlemmings} {et~al.}(2011){Vlemmings}, {Humphreys}, \&
  {Franco-Hern{\'a}ndez}}]{vh11}
{Vlemmings}, W.~H.~T., {Humphreys}, E.~M.~L., \& {Franco-Hern{\'a}ndez}, R.
  2011, \apj, 728, 149

\bibitem[{{Wolleben} \& {Reich}(2004)}]{wr04}
{Wolleben}, M. \& {Reich}, W. 2004, \aap, 427, 537

\end{thebibliography}
\end{document}